\documentclass[twocolumn,10 pt,superscriptaddress,amsmath,amssymb,showpacs]{revtex4}
\usepackage{graphicx}
\begin{document}
\title{On voltage-current characteristics and critical current in Bi-2212}
\author{A. Pautrat}
\email[corresponding author: ]{ alain.pautrat@ensicaen.fr}
\author{Ch. Simon}\author{J. Scola}\author{C. Goupil}
\affiliation{CRISMAT/ENSI-Caen, UMR 6508 du CNRS,6 Bd Marechal Juin, 14050 Caen, France.}
\author{A. Ruyter}\author{L. Ammor}
\affiliation{LEMA-Universit$\acute{e}$ de Tours ­ UMR CNRS/CEA 6157 - 37000 TOURS ­
France.}
\author{P. Thopart}\author{D. Plessis}
\affiliation{CEA/DMAT ­  B.P. 16 ­ Le Ripault - 37260 MONTS ­ France.}

\begin{abstract}
V(I) characteristics have been performed in a monocrystalline microbridge of
$Bi[Pb]-2212$. The vortex phase diagram has been greatly investigated. Linear but
non-ohmic Voltage(Current) (V(I)) curves with well defined critical current have been
observed. A departure from this behavior is observed near the peak effect where an out of
equilibrium high threshold current can be stabilized. At high temperature, the critical
current persists in the "liquid" state despite the dissipation at the lowest bias. Some
implications of these results are discussed. In particular, it is proposed that the
surface disorder, rather than the bulk disorder, is responsible for the vortex pinning in
this sample.
\end{abstract}
\newpage
\maketitle

The generic vortex phase diagram in cuprates is now described in terms of into different
thermodynamic phases of vortex \cite{GB}. The main idea is that an ordered Vortex Lattice
(VL) is present at low field and low temperature, and that it develops into a phase
possessing a degraded order when the thermal fluctuations or the static disorder are
increased. This corresponds respectively to (ordered solid/liquid) and (ordered
solid/disordered solid) transitions. The strong experimental facts which have justified
these ideas are the disappearance of the vortex pinning in the "liquid" phase and the
increase of this vortex pinning (the peak effect) when crossing the disordering
transition. The highly anisotropic cuprate Bi-2212 can be taken as a representative
sample where these three states can appear. Nevertheless, recent experiments suggest to
qualify this point of view. Indeed, it has been found that both the high field and the
high temperature properties can be interpreted with the same state of vortex matter,
meaning that no difference between them should be presupposed \cite{kees,zeldov}. Another
striking result is that VL translational order is not a good order parameter to
characterize the first order transition \cite{meng}, whereas one could have expected the
contrary for a genuine melting. From an experimental point of view, the VL behavior in
Bi-2212 has been tackled at low field by magneto-optic imaging or local ac probes
\cite{kees,zeldov}, which are sensitive to pinning induced screening currents on the
surface of the samples. If the magnetic field is increased, the limited resolution of the
above-cited experiments strongly restricts their ability to give information on the VL
pinning and dynamical properties. One complementary and easily understandable experiment
is the measure of a voltage versus current (V(I)) curve. The critical current $I_c$,
which gives the pinning ability of the medium, can be extracted without the need of
complicated and possibly unjustified assumptions. Furthermore, it has been shown by
numerous theories and simulations \cite{marc, giam, olson} that the nature of the VL
should govern the variation of its velocity as function of a bulk training force. To the
extent that this force is directly given by the amount of applied transport current,
these signatures are directly expected in the V(I) curves. For example, there is some
consensus concerning the dynamical properties of a VL \cite{GB}. They are expected to be
dominated by hopping over barriers at low currents, with a transition to a free flow and
Ohmic-like regime at high current. Fig.1 is a schematic representation of the different
V(I) curves which can be expected for the different VL states, namely ordered state
(Bragg Glass) and disordered states (vortex Glass and vortex liquid). We note that if
numerical simulations of $\textit{velocity versus force}$ curves are extremely numerous
in the literature, very few direct measurements have been performed in cuprates to see if
the main expectations compare well with the experiments. In the present experiment, we
are thus principally interested in the qualitative differences (if any) between the
functional form of the V(I) curves, for different locations (B,T) in the phase diagram.

The measurement of a V(I) curve is conceptually a very simple method and is currently
used as a useful probe of vortex properties in low T$_{c}$ materials. On the contrary, it
was unemployed for high T$_{c}$'s, except using very low current densities. The reason is
practical: quasi-perfect contacts are usual in low T$_c$ materials ($R_{contacts}
\lesssim 0.001\Omega$) but not in cuprates ($R_{contacts} \gtrsim$ 1-2 $\Omega$). It is
thus difficult to avoid any overheating in these rather resistive contacts as soon as the
current exceeds tens of milliAmperes. Typically, under normal Helium atmosphere and
standard experimental conditions, I$\thickapprox$100 mA into a $1 \Omega$ metallic
contact ($\lesssim$ mm$^{2}$) is compatible with a local increase of the temperature of
about 0.1 K. This strongly restricts the value of the "safe" transport current.
Extrapolating to a typical sample size (W=1 $mm$, t= 100 $\mu m$, W=width, t=thickness),
the V(I) curves are thus difficult to perform as soon as $J \gtrsim 10^2 A/cm^2$ ($i
\gtrsim 0.5 A/cm$ if expressed in surface current units). Since the critical current
should be at least less than these values, only the depinning onset close to the high
temperature first order transition can be reliably measured. It is worth noting that
similarly low current densities are usually not sufficient to overcome unavoidable
critical current inhomogeneities at the onset of vortex depinning, even in pure
superconducting metals. In this regime, only parts of the VL are moving. To reach the
$\textit{flux-flow}$ regime where the whole VL is in motion, it is necessary to increase
the value of the injected current, say $(I-I_c)$ reasonably larger than $I_c$. One of the
possibilities which allows to overcome the overheating problems is the use of fast
current pulses. This technique was employed in \cite{sas}. Both stable linear and
metastable V(I) curves with a S-shape were measured. Since the pulse of current injection
is very fast (about 100 $\mu$s), a normal skin effect can affect the preceding results
and it is not completely clear if they can be taken as representative of the steady state
of the moving VL. A confirmation of this experiment, using a continuous transport
current, appears thus necessary. For that, we have studied monocrystalline bridges of 100
or 200 $\mu$m of width. This ensures a good homogeneity of the current injection. Above
all, for a same current I, a largest current density ($I/tW=J$ or $I/2W=i$) is obtained
when using a microbridge rather than a bulk sample. This let the possibility of working
in the $\textit{flux-flow}$ regime without the overheating at the resistive contact pads,
even at low temperature.

In this experiment, we would like to address simple experimental facts to the following
questions:

- What are the fundamental differences, if any, in the shapes of the V(I) curves for the
different VL states? - Is it possible to explain the critical current values?

 The samples used in this study are slightly Pb doped single
crystals of the Bi-2212 family ($Bi_{1.8}Pb_{0.2}Sr_2CaCu{_2}O_{8+\delta}$). They were
grown by the self-flux technique as previously described \cite{ruyt}. Each cleaved single
crystal was laser tailored in the form of a microbridge with a controlled pattern of
(W=200 * L=400) $\mu m^2$ or (W=100 * L=400) $\mu m^2$ (Fig.2). The thickness is about
100 $\mu m$. The crystal was after annealed under a controlled pure oxygen gas flow and
is in the slightly overdoped regime ($T_c$= 79.5 K). Low resistance electrical contacts
($\leq 1 \Omega$) were made by bonding gold wires with silver epoxy. The DC transport
measurements were performed using a standard four probe method (cryostat Quantum Design
with a 9T horizontal coil, external current source Adret and nanovolmeter Keithley). The
results presented here correspond to the microbridge with W= 200 $\mu m$, otherwise it
will be specified in the text.

 RESULTS IN THE LOW TEMPERATURE REGIME

 We have first performed V(I) curves at low temperature (T = 5K) in order to
 minimize the effect of the thermal fluctuations. Let us discuss the
results for high magnetic field values (1T $\geq$ B $\leq$ 9T, Fig. 3). The V(I) curves
are very similar to those observed for a conventional vortex lattice, as it can be
measured in low $T_{c}$ metals or alloys. In particular, they present the usual form $V=
R_{ff} (I-I_c)$ as soon as I is slightly higher than the critical current $I_c$ ($R_{ff}$
is the $\textit{flux-flow}$ resistance of the VL). There is no evidence of an ohmic
regime at low applied current and the depinning is rather stiff. Furthermore, we have
tried to compare the effects of a Field Cooling (FC) and of a Zero Field Cooling (ZFC),
or of a FC under different cooling rates. Always the same dissipation has been measured
in the time scale of our experiment. In particular, no aging effect is observed on the
critical current. This is not in agreement with a glassy nature of the VL governing
transport properties.

When the magnetic field is decreased, a different behavior is observed in a restricted
region of the phase diagram ($0.05 T \leq B \leq 0.2 T$). If the VL is prepared after a
FC, the V(I) curve exhibits a S-shape with a high threshold current $I_{high}$
\cite{notabene}, but only for the first increase of the current. After this initial ramp,
a reproducible $I_c < I_{high}$ is always measured (Fig. 4). This has been previously
observed in the pulsed-current experiments, and this state with a "high threshold
current" has been evidenced as a metastable state with a very long relaxation time
\cite{sas,portier}. Our measurement using a dc current proves that the observation of
this state is not due to the kind of stimulation used. This observation is also in
agreement with the observed supercooling of a high critical current state using a
time-resolved local induction measurement \cite{kees}. We observe a hierarchy in the
accessible threshold currents. Depending on the exact preparation of the FC state
(cooling rate, value of the initial field, rate of the current injection ramp), numerous
threshold values are accessible between $I_{high}$ and $I_c$. It is clear that in such a
regime where it is easy to lock an $\textit{out of equilibrium}$ state, most of the
measurements will give transient and spurious relaxation effects if the field is ramped
or quenched as it is done in magnetization measurements. We find some traces of this
metastability up to about 1T, but the most obvious effects are restricted in the region
(0.05-0.2T) (Fig.5). For $I
> I_{high}$, the V(I) curve is observed disrupted. If the width of the microbridge is narrower, multiple steps in the
current induced moving state can even be observed (inset of fig. 4). Even small (100 $\mu
m$), this width is much larger than any superconducting length. It is thus likely that
these structures in the V(I) curves can not be explained by phase slippage processes as
observed in one dimensional superconductors \cite{sko}. This mimics rather the voltage
steps which have been observed in the current-induced resistive state of type I
superconductor \cite{step} and can be the counterpart of a coexistence between two states
possessing different critical currents and comparable spatial extensions. This dynamic is
close to what is observed in a current carrying inhomogeneous superconductor. It can not
be exclude that the high values of the threshold current are responsible for local
heating in the interfaces between the domains and favor the formation of resistive
"electrothermal" structures \cite{mints}. Concerning the metastable V(I) curves, the peak
effect in the critical current, the coexistence of two VL states, the same kind of
behavior is currently observed in $2H-NbSe_2$ \cite{shobo}. The strong difference is that
the peak effect and the associated metastable effects appear close to $B_{c2}$ in
NbSe$_2$ but is here restricted to a very low field value. Since the applied temperature
is very similar in both experiments, the explanation of this field value difference has
to be found in the large difference between the electronic anisotropies. For a field
lower than about 0.05T, we do not observe any hysteresis within the V(I) curves. This is
summarized in fig. 5 where $I_{c}$ versus $B$ is shown.

One has to remark that the variation of $I_c(B)$, if one excepts the small low field part
where metastability takes place, looks like what is measured in soft low T$_c$ materials.
To some extent, one can speculate that the same pinning mechanism is acting without
involving a different VL phase. Qualitatively, the functional form of $I_c(B)$ is very
close to the one of the reversible magnetization of a high $\kappa$ anisotropic
superconductor \cite{patrice}, meaning that $I_c$ is linked to the weight of the
diamagnetic screening currents. This has to be brought close to the linear V(I) curves
that we measured. We will return to this point later.

HIGH TEMPERATURE RESULTS: THE "LIQUID" STATE

Let us now discuss the V(I) curves obtained at high temperature. The temperature T was
fixed at 50 K and the magnetic field B was varied from 0.001T to 9T, in order to cross
the transition between the so-called "solid" and "liquid" vortex phases. In Bi-2212,
thermodynamical consistency required by the respect of the Clausius-Clapeyron relations
\cite{marcenat} can not be proved, because up to now no transition can be detected using
specific heat measurements. The small step usually observed in the magnetization
\cite{kaik}, added to the reasonable idea that the same physics is acting both in Bi-2212
and in YBaCuO, can be taken as a good indication that this transition is of first order.
The appearance of an approximate linear resistivity when the sample is probed at very low
current density is also usually taken as a good criteria. The underlying ideas are that a
liquid state of 2D vortices can not be pinned and that the resistive properties are close
to that of a metal (Ohmic regime). Using the same criterion, the transition would be
located at about 0.08 T (fig. 6), in agreement with values currently reported in the
literature (0.065 T in \cite{kaik} for a Bi-2212 with an equivalent doping).
Nevertheless, we observe that for magnetic fields largely higher, in a large part of what
is identified as the "liquid" state, the V(I) curves are clearly non ohmic (fig. 7). A
non linear response in this resistive state has already been observed. It has been
interpreted as the feature of a highly viscous liquid of vortices \cite{tsuboi}, or by
the edge effect of surface barriers (SB) \cite{zeldov2}. Roughly speaking, SB are
expected to be negligible for fields $B\geq B_{c1} \kappa / ln (\kappa) \approx 20
B_{c1}$ for high ${\kappa}$ superconductor (${\kappa}$ = 100) with an ideal flat surface
\cite{barriere}. Any real sample also possesses surface irregularities which facilitate
vortex nucleation and decrease this value. We estimate that, taken $B_{c1}\leq 300G$, SB
effects can be neglected when a magnetic field of several Teslas is applied. We note also
that other experiments performed using the Corbino geometry do not evidence any effect of
SB at high temperature \cite{mazilu}. We have also checked that a decrease of the sample
width decreases $I_c$, confirming that neglecting SB is reasonable, at least as far as
critical current properties at high field values are involved.
 We stress on the following points: the high current part of the V(I)
curves is linear and its extrapolation never goes to zero (fig. 7), at least for B $\leq$
9 T. In other terms, putting aside the rounded dissipation onset, V(I) curves can be
expressed as $V = R_{ff} (I-I_c)$ as it was observed at low temperature. It can not be
explained by a non linear mechanism in which a depinning energy would be a function of
the driving force required to overcome barriers, because the high current regime is
linear but does not reach an asymptotic Ohmic regime ($V = R.I$). This demonstrates the
existence of a real critical current $0 \leq I_c \leq
 10 mA$ in this high temperature state. This critical current exhibits a field
dependence that compares well with the one obtained at low temperature. It is thus
reasonable to think that the same pinning mechanism occurs at low and high temperatures.
The difference in the low dissipation level can be understandable in terms of an additive
process which appears at high temperature, and a thermally activated process which
assists the depinning is a natural candidate. This was largely studied in the literature
\cite{GB}. This defines a threshold current I* above which a small dissipation appears
but pinning continues to exist (Fig.6).

DISCUSSION

From a theoretical point of view, vortex lattice depinning is generally described like a
critical phenomena: the bulk depinning of an elastic system in a random media. Driven
states are described with an overdamped dynamical equation and it is expected that
without thermal activation the velocity $v$ just above the depinning scales like
$(F-F_c)^{\beta}$  where $F$ is the applied force, $F_c$ the bulk pinning force and
$\beta$ is the critical exponent \cite{GB, bulk}. The usual analysis supposes that
$(F-F_c)^{\beta}$ identifies with $(J-J_c)^{\beta}$, $J$ being a current bulk density. At
high velocity, disorder is found not relevant that leads to a velocity $v \alpha F$.
Experimentally, this should correspond to an Ohmic regime. Clearly, the experimental
results are different. Even without performing a detailed analysis, one can realize that
the V(I) curves at the depinning onset are never observed convex. $\beta$ is never lower
than 1 even at low temperature where the vortex Creep driven by thermal activation is
negligible. One excludes from this discussion the peculiar case of the peak effect where
a convex part is effectively observed but is very likely due to the macroscopic
inhomogeneity of the state. It is more important to remark that, when $I \gg I_c$, the
velocity is found to vary like $(I-I_c)$ and not like $I$. This means that the pinning
force felt by the VL remains constant even at "high velocity", in apparent contradiction
with an explanation in terms of a bulk depinning. One could object that the applied
current in our experiment is not sufficiently high to reach the predicted Ohmic-like
regime. This problem can never be strictly resolved because Joule heating is always a
limitation when increasing the current in a transport experiment. Nevertheless, one can
refer to low T$_c$ materials where the experimental situation is much more attractive.
Under very controlled temperature, it is possible to verify $V \alpha (I-I_c)$ up to at
least $I \geq 30 I_c$. This is thus the generic shape of a V(I) curve in a type II
superconductor, and the present experiment shows that the same dissipation mechanisms are
acting in Bi-2212 samples. It has been also verified by inelastic neutron scattering
that, in this regime, the VL is moving freely as a whole \cite{echo}. This latter result
means that the bulk disorder can be estimated to be no more effective. At the same time,
the former result shows that the velocity does not reach the asymptotic regime where the
velocity should vary linearly with the total current. To be coherent with the above
mentioned theories \cite{bulk}, one possible solution is to replace $\textit{by hand}$
the applied force $F$ (resp $I$) that acts against bulk disorder by $(F-F_c)$ (resp
$(I-I_c)$) (Fig. 9). Physically, this solution appears in the case of vortex pinning by
the surface roughness \cite{hocquet}. The main idea is that the surface disorder (a quite
standard surface roughness) allows, thanks to boundary conditions, for the flow of non
dissipative superficial current ($i_c (A/cm)=I_c / 2W$). This is only when all the
superficial non dissipative paths are exhausted, precisely when $I \geq I_c$, that the
excess of current $(I-I_c)$ penetrates the bulk. To some extent, a bulk force makes then
sense and a "bulk depinning" can occurs, but involving only the over critical part of the
applied current. The bulk disorder can be averaged by the motion without affecting the
main critical current which reflects the surface pinning ability, explaining the
experimental shape of the V(I) curves. As an further indication that this mechanism can
occur in the Bi-2212 samples, quantitative expressions for $I_c(B)$ are predicted and can
be checked. The case of very anisotropic samples is specially interesting, because it is
predicted that for not too low magnetic field values and for a moderate surface
roughness, the surface critical current $i_c$ may become independent of the surface
quality and depends then solely on the parameters of the condensate \cite{patrice}. For
clarity, we restrict the comparison to the high field values in order to use the
Abrikosov limiting expressions \cite{patrice}:

\begin{equation}
i_c = \frac{B_{c2}}{2 \mu_0 \beta_{A} \gamma \kappa_{2//c}(T)^{2}}
(1-(B/B_{c2})^{2/3})^{3/2} \label{Ic}
\end{equation}

Here, $\kappa_{2//c}(T)$ is the generalized Ginzburg-Landau parameter, $\gamma$ the
electronic anisotropy and $B_{c2}$ the second critical field. For the small temperature
dependence of $\kappa_2$, we use the microscopic GL derivation \cite{kappa}. This
expression describes rather well the data (Fig.10), with the use of a restricted number
of parameters. $\gamma=60$ was obtained from angular resolved resistivity measurements
using scaling arguments \cite{scaling}. With $\kappa_2=\frac{\lambda}{\xi}$,
$B_{c2}=\frac{\phi_0}{2\pi\xi^2}$, we find $\xi_{(0)}= 3.3 nm, \lambda_{(0)}= 2080 nm$
consistent with the slightly overdoped regime \cite {param}). The approximate field
dependence of the critical current in a restricted field range is an indication but can
not be taken as a proof of the validity of the model. Other field dependence would be
acceptable. In particular, in a log-log representation, it is reasonable to estimate that
$I_c$ varies like $B^{-0.5}$, as it is often observed (\cite{sas} and references herein)
and interpreted as a 2D strong bulk pinning regime. A possible limitation of this latter
analysis is the relevance of the strong bulk pinning centers in clean single crystals. It
is more important to note that the absolute value of the critical current is the one
expected if only the surface contributes to pinning. To compare with other experiments,
we measured $i_c$ $\approx$ 0.1-0.5 A/cm at (T= 5K, B= 9 to 1T) in our microbridge, and
0.4-1 A/cm was measured in \cite{sas}. These values seem representative of Bi-2212
crystals. Now it can be objected that the current distribution in Bi-2212 can be strongly
influenced by the very large electronic anisotropy \cite{pethes}. This leads to a very
small current penetration in the normal and in the superconducting state and this has
$\textit{a priori}$ no link with any pinning mechanism. It remains that if a current
flowing under the surface is non dissipative up to a critical value, a surface pinning
mechanism should apply. Finally, it seems that there is no need to involve bulk disorder
in order to explain the critical current and the shape of the V(I) curves at least
outside the peak effect region. Other experiments will soon be performed to evaluate more
precisely the validity of this hypothesis and the specific case of the peak effect will
be discussed in another paper.

 It is clear that this interpretation of the VL pinning in Bi-2212 is at odds with the
 currently accepted view that the pinning is driven by the bulk disorder (collective bulk pinning).
 We need to add some comments to replace this result in the context of recent experiments.

In Bi-2212, the peak effect should separate a high field disordered phase with glassy
properties the low field ordered VL. The VL, corresponding to the V(I) curves of the
fig.3, is expected to be in this glassy state, whereas we observe a conventional VL
dynamic rather than a glassy dynamic in its proper sense. The experimental proof of a
disordered high field state in Bi-2212 is also not completely obvious. Small Angle
Neutron Scattering experiments can effectively be interpreted in terms of a VL
disordering transition at very low field \cite{bob}. This interpretation relies in the
strong decrease of the diffracted intensity when the density of vortices increases,
whereas a simple London model which disregards the finite size of the vortex core
predicts only a smooth decrease due to geometrical factors. Nevertheless, recent
theoretical progresses allow to calculate the microscopic field distribution and its
Fourier components are found to be strongly field dependent even at low field (for high
kappa and/or anisotropic superconductors) \cite{brandt}, meaning that involving a
transition in the VL may be not necessary. This can explain why other experiments with a
largest intensity are showing that a well ordered VL with good Bragg peaks survives at
higher fields \cite{ted}. We note also that recent experiments suggest that the VL order
is not relevant for the nature of the "melting" transition in Bi-2212 \cite{meng},
meaning that the order parameter
 of this transition is not linked to the topology of the VL. Experimental data suggest
  that the line of the peak effect in the phase diagram is
the continuation of the first order line into the low-temperature \cite{kaik}. This
implies that the same state of the VL is present at low and high temperature. Our
experiment suggests that the pinning of the vortex lattice is also of the same nature,
even if thermal fluctuations are responsible for a low dissipation background at high
temperature.
 Finally, it has been recently proposed
 that the renormalization of the (non-local) line tension by thermal fluctuations for large wave vectors is at the origin of the first order
 transition \cite{colson}.
As the line tension is the controlling parameter for the surface pinning, this offers a
 scenario more compatible with our results than the occurrence of a genuine melting transition.
Furthermore, even if the bulk condensate is strongly fluctuating with eventually
disrupted vortices, a small dissipation-less current can persist under the surfaces. A
well known example is the case of surface superconductivity where a critical current does
exist in the surface sheath even if the bulk develops into its normal state.

 To conclude, the use of monocrystalline microbridges of Bi-2212 allowed us to measure
 V(I) curves in different parts of the VL phase diagram. They appear conventional and
a clear critical current can be defined by extrapolating the high current linear part. A
metastable high threshold state can be stabilized by Field Cooling. This corresponds to
the peak effect region and mimics what is observed much closer to $B_{C2}$ in NbSe$_2$.
The values of the measured critical currents and the shapes of the V(I) curves suggest
that the pinning by the surface roughness can not be ignored for understanding the vortex
lattice dynamic in Bi-2212. We observe that this superficial critical current survives in
a large part of the so-called liquid state. Finally, the overall picture suggests that
the VL pinning and dynamic in Bi-2212 are more similar to what is currently observed in
low $T_c$ materials than it is often suggested.

Acknowledgments: Joseph Scola acknowledges support from "la r$\acute{e}$gion basse
Normandie".

\newpage

\begin{figure*}[tpb]
\centering\includegraphics*[width=6cm]{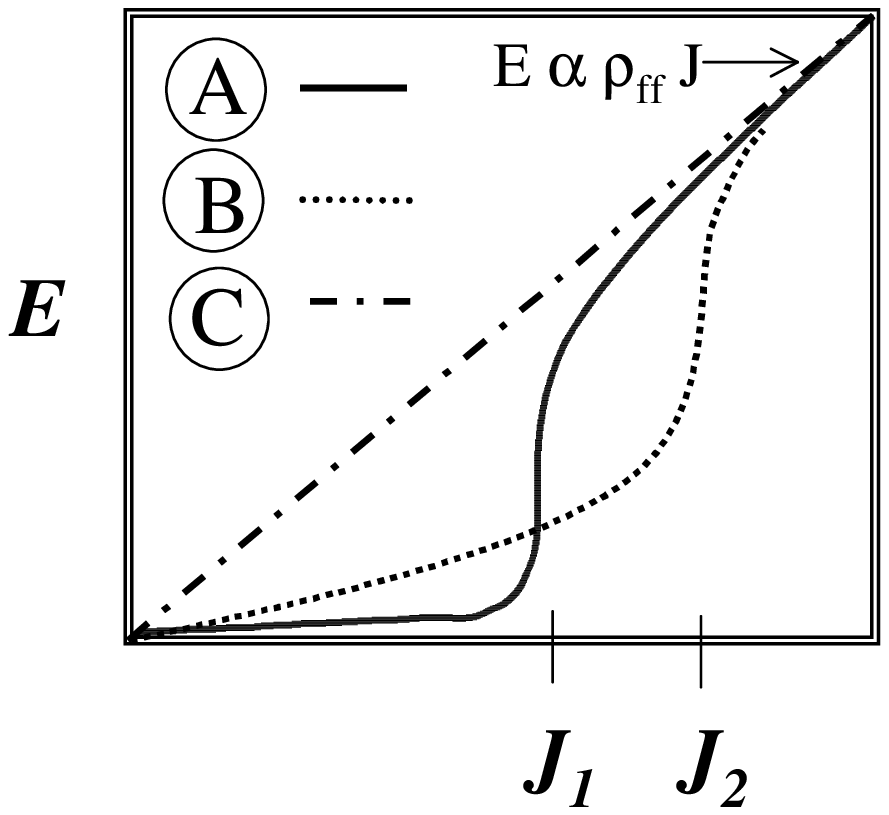}
 \caption{V(I) curves expected for different phases of the VL
(A:Quasi-Lattice (Bragg Glass), B:Glass, C:Liquid). $J_1$ represents the depinning
current in the phase A, $J_2$ in the phase B.} \label{f.1}
\end{figure*}
\newpage
\begin{figure*}[tpb]
\centering \includegraphics*[width=6cm]{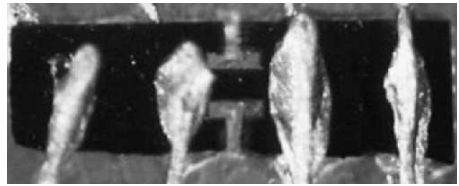} \caption{The monocrystalline
microbridge (200*400 $\mu m^2$) of Bi-2212.} \label{f.2}
\end{figure*}
\newpage
\begin{figure*}[tpb]
\centering\includegraphics*[width=6cm]{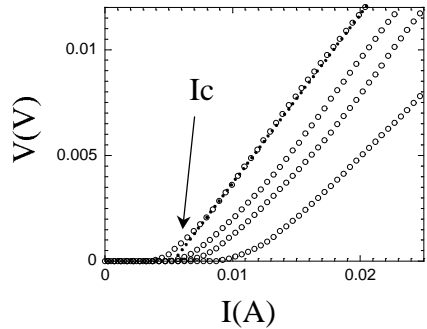}
 \caption{Reversible V(I) curves (T = 5 K, B = 1, 3,
5, 9 T from the right to the left). The dashed and straight line is a guide for the eyes
and evidences the $\it{flux-flow}$ regime.
} \label{f.3}
\end{figure*}
\newpage
\begin{figure*}[tpb]
\centering \includegraphics*[width=6cm]{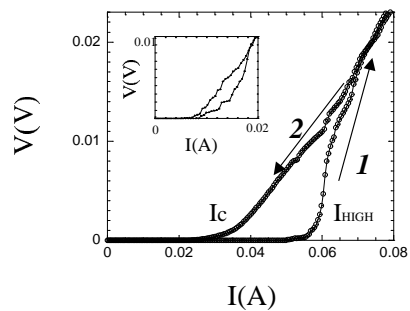}
 \caption{Hysteretic V(I) curve (T = 5 K, B = 0.15 T after a FC).1/ The first increase of current defines $I_{high}$ and 2/ The following decrease of
the current defines $I_c$ . In the inset is shown the same kind of curve in the
microbridge with a smallest width (T = 5 K, B = 0.05 T). Note the steps in the V(I)
curve.} \label{f.4}
\end{figure*}
\newpage
\begin{figure*}[tpb]
\centering \includegraphics*[width=6cm]{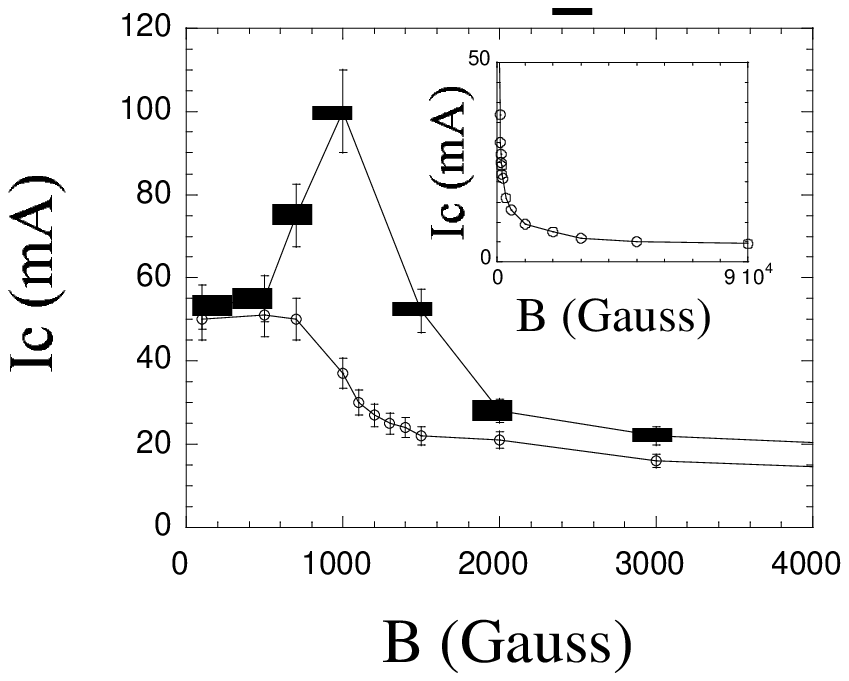} \caption{Peak effect in the critical
current (T=5K. Black points, $I_{high}$ and empty points, $I_c$). In the inset is shown
the variation of $I_c$ at high field.} \label{f.5}
\end{figure*}
\newpage
\begin{figure*}[tpb]
\centering \includegraphics*[width=6cm]{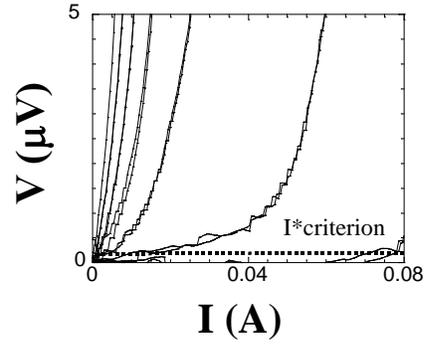} \caption{Low dissipation level of the
V(I) curves (T=50K, from the left to the right: B= 0.14 T, 0.12 T, 0.1 T, 0.08 T, 0.06 T,
0.04 T, 0.02 T). As shown in the figure, $I_{c}^{*}$ is defined using a criterion of 0.2
$\mu V$ and is nearly zero for B $\gtrsim$ 0.08 T.} \label{f.6}
\end{figure*}
\newpage
\begin{figure*}[tpb]
\centering \includegraphics*[width=6cm]{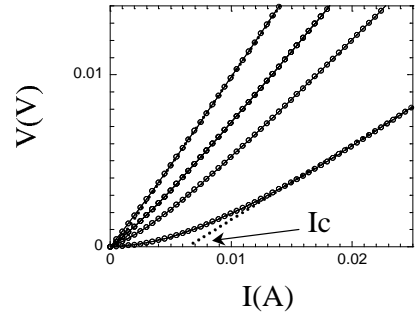} \caption{Reversible V(I) curves (T =
50 K, B = 1, 3, 5, 9 T from the right to the left). The dashed and straight line is a
guide for the eyes. This linear extrapolation defines $I_c$. Note that the regime is
clearly non ohmic up to at least B = 9 T.} \label{f.7}
\end{figure*}
\newpage
\begin{figure*}[tpb]
\centering \includegraphics*[width=6cm]{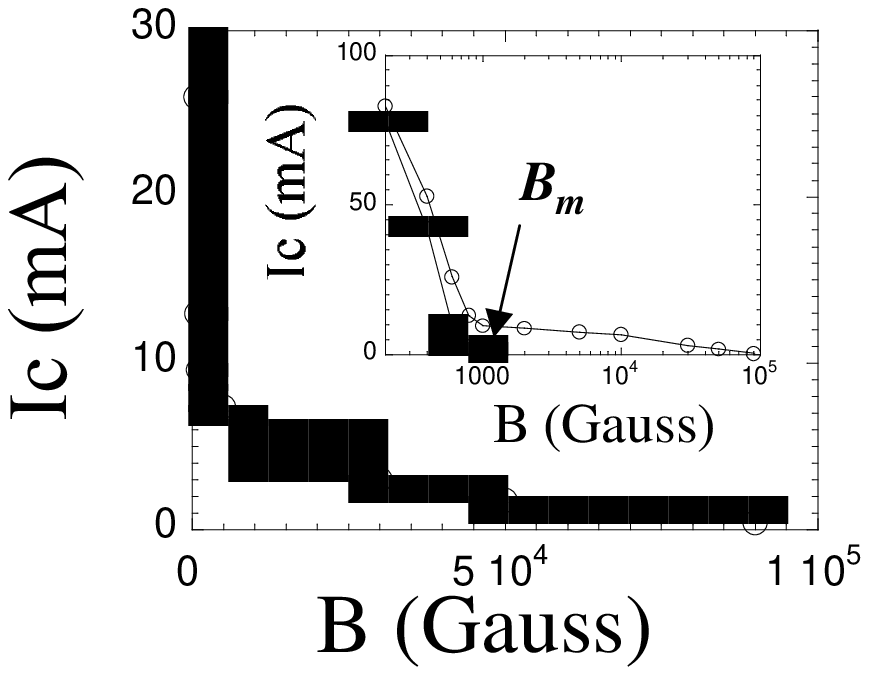}
 \caption{$I_c (B)$ at T= 50 K. In the inset is shown the low field
part using a semi-log representation (black point, $I^{*}$ and empty point, $I_c$). Note
the disappearance of $I^{*}$ at $B_m \leq$ 0.08 T whereas $I_c$ keeps a non zero value.}
\label{f.8}
\end{figure*}
\newpage
\begin{figure*}[tpb]
\centering \includegraphics*[width=6cm]{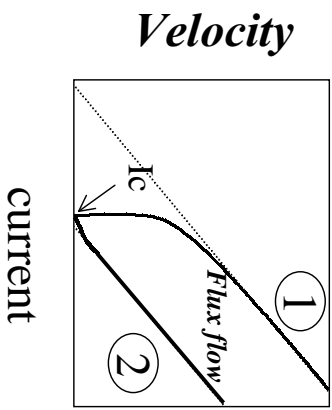}\caption{Differences between the V(I)
curves corresponding to (1) a bulk depinning involving all the current ($V \alpha I$ in
flux-flow) and (2) a surface depinning with Ic (non resistive) remaining close to the
surface at high velocity ($V \alpha (I-I_c)$ in flux-flow).} \label{f.9}.
\end{figure*}
\newpage
\begin{figure*}[tpb]
\centering \includegraphics*[width=6cm]{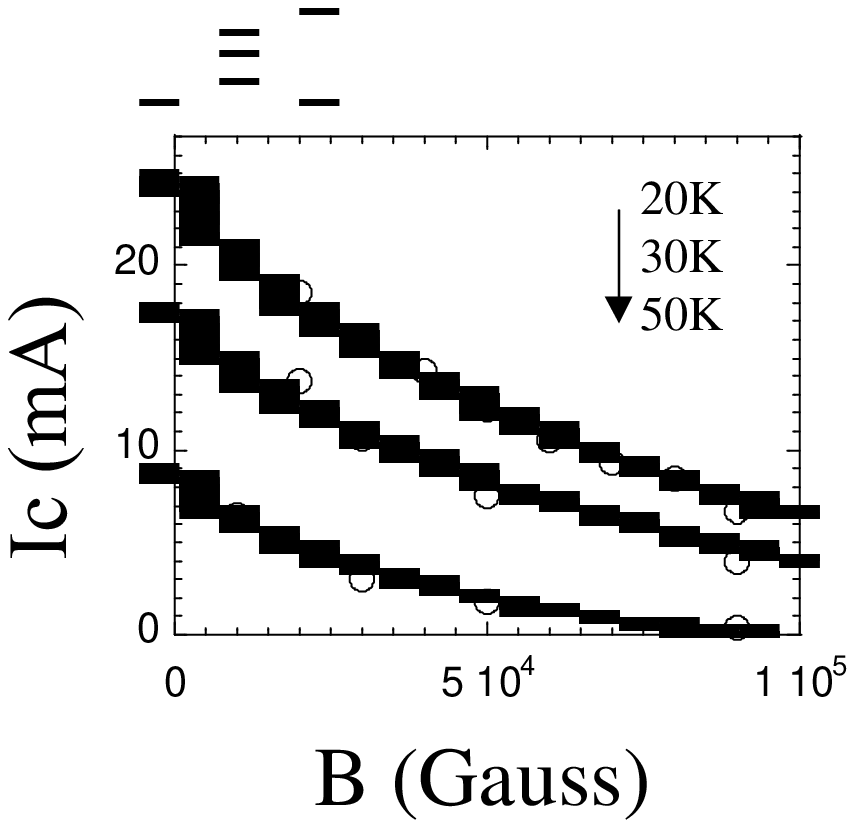}
 \caption{$I_c(B)$ at T=20, 30 and 50 K for the high magnetic field values. The
solid line is a fit using the formula (1) as it is explained in the text.} \label{f.10}.
\end{figure*}


\begin{references}
\bibitem{notabene} Since $I_{high}$ is not reflecting the equilibrium state, we use the
term threshold current, and we limit the use of the term critical current for the
depinning current which is not history dependent.
\end{references}
\end{document}